\documentstyle[twocolumn,aps,epsfig,floats]{revtex}
\begin{document}

\twocolumn[\hsize\textwidth\columnwidth\hsize\csname %
@twocolumnfalse\endcsname

\title{Effect of phase fluctuations on INS and NMR experiments in the pseudo-gap 
regime of the underdoped
cuprates}
\author{Harry Westfahl Jr.$^1$ and Dirk K. Morr$^{1,2}$ }
\address{$^1$ University of Illinois at Urbana-Champaign, Loomis Laboratory of
Physics, 1110 W. Green St., Urbana, IL 61801 \\ $^2$ Theoretical
Division, Los Alamos National Laboratory, Los Alamos, NM 87545 }
\date{\today}
\maketitle
\draft

\begin{abstract}
We present a theory for inelastic neutron scattering (INS) and
nuclear magnetic resonance (NMR) experiments in the pseudo-gap
regime of the underdoped high-$T_c$ cuprates. We show that
superconducting phase fluctuations greatly affect the temperature
and frequency dependence of the spin-susceptibility, $\chi''$,
probed by both experimental techniques. This result explains the appearance
of a {\it resonance peak}, observed in INS experiments, below a
temperature $T_0 > T_c$. In the same temperature
regime, we find that the $^{63}$Cu spin-lattice relaxation rate,
$1/T_1$, measured in NMR experiments, is suppressed. Our results
are in qualitative agreement with the available experimental data.
\end{abstract}
\pacs{PACS numbers:74.25.-q,74.25.Ha,74.25.Jb,74.40.+k}

]

Over the last few years intensive research has focused on the
origin of the pseudo-gap region in the underdoped high-$T_c$
cuprates \cite{Eme95,Fra98,Kwo98,Chu98,Ges97,Zha97,Lee97}. This
part of the phase diagram, below a characteristic temperature $T_{*}>T_{c}$,
is characterized by a suppression of the low-frequency quasi-particle spectral
density, as observed by angle-resolved photo-emission (ARPES) \cite{arpes} and
scanning tunneling spectroscopy (STS) experiments \cite{sts}. For the same
compounds, inelastic neutron scattering (INS) experiments have revealed a
sharp magnetic mode, the {\it resonance peak}, below $T_*$, in
contrast to the optimally doped cuprates, where it only appears
below $T_c$ \cite{Dai}. Moreover, nuclear magnetic resonance (NMR)
experiments find a strong decrease of the $^{63}$Cu spin-lattice
relaxation rate, $ 1/T_{1} $, below $T_{*}$  \cite{Tak91,Ish98}.

These experimental observations put tight restrictions on the
proposed theoretical scenarios for the pseudogap ascribing it to
spin-charge separation \cite{Lee97}, SO(5) symmetry \cite{Zha97},
condensation of performed pairs \cite{Ges97} and spin-fluctuations
\cite{Chu98}. Emery and Kivelson (EK) \cite{Eme95} proposed that,
due to the small superfluid density of the underdoped cuprates,
thermal fluctuations in the phase of the superconducting (SC)
order parameter destroy the long-range phase coherence in the
pseudo-gap regime, while preserving a finite local amplitude of
the order parameter, $|\Delta(r)|$. In this communication, we
argue that the presence of phase fluctuations provides an
explanation for the results of INS and NMR experiments discussed
above. We show that these fluctuations greatly affect the
temperature and frequency dependence of the spin-susceptibility,
$\chi''$, probed by both experimental techniques. Support for the
existence of phase fluctuations comes from recent high frequency
transport experiments by Corson {\it et al.} \cite{Cor99}. They
demonstrated that the SC transition in underdoped
Bi$_2$Sr$_2$CaCu$_2$O$_{8-\delta}$ (Bi-2212) is of the
Kosterlitz-Thouless (KT) type, where at $T_c=74K$ the unbinding of
thermally excited vortex-anti-vortex pairs destroys the long-range
phase coherence. However, they also concluded that $|\Delta(r)|$
vanishes at a temperature $T_0 \sim 100K$, while the onset
temperature, $T_*$, for the pseudo-gap regime is much higher. For
this reason we will focus our analysis on the region $T_c<T<T_0$.

The starting point for our calculations is the mean-field BCS
Hamiltonian in which the phase, $\theta(r)$, of the
superconducting order parameter, $\Delta(r)=|\Delta(r)| \,
e^{i\theta(r)}$, varies on the scale of the phase coherence
length, $\xi_{\theta}$. Such a nonuniform phase $\theta({\bf r})$
is treated via a gauge transformation \cite{Fra98,Kwo98}
\begin{equation}
\Psi^{\dagger}=e^{i\theta(r) /2}c^{\dagger }
\label{GT}
\end{equation}
where $c^{\dagger }$ is the creation operator of the
original  electrons. This transformation induces a coupling of the
$\Psi$-fermions to a local superfluid flow
$\bf{v}_{s}(\bf{r})=\nabla \theta(\bf{r})/2m$,
(we set $\hbar=1$) whose thermodynamic properties are determined by the 2D-XY
Hamiltonian \begin{equation}
{{\cal H}_{XY}\over k_BT}={K_{0}(T)\over 2} \,
\int d^2\bf{r}\, |\nabla \theta (\bf{r})|^2 \ ,
\label{Hxy}
\end{equation}
where $K_{0}(T)=n_{s}(T)/(4mk_B T)$ is the ``bare'' phase
stiffness and $n_{s}(T)$ is the 2D superfluid density per CuO$_2$
layer which for a d-wave superconductor is given by $n_{s}(T) =
n_{s}(0)(1-T/T_0)$, where $T_0$ is the BCS mean field temperature
\cite{Lee97}. In order to compute $\chi''$ in the presence of
phase fluctuations we first compute it for a given configuration
of $\bf{v}_{s}(\bf{r})$ and subsequently perform a thermodynamic
average over the ensemble specified by Eq.(\ref{Hxy}). Our
approach is similar to the one adopted by Franz and Millis (FM)
\cite{Fra98} who computed the single particle Green's function,
$G(k,\omega)$, in the pseudo-gap regime. They showed that the
quantity which determines the ensemble average is the correlator
$W = m^2 v_F^2 \langle {\bf v}^2_{s} \rangle/2$, whose temperature
dependence they extracted from fits to ARPES and STS experiments.
In the following we show that $W(T)$ can also be obtained from the
experiments by Corson {\it et al.}~\cite{Cor99}

\begin{figure} [t]
\begin{center}
\leavevmode
\epsfxsize=7.5cm
\epsffile{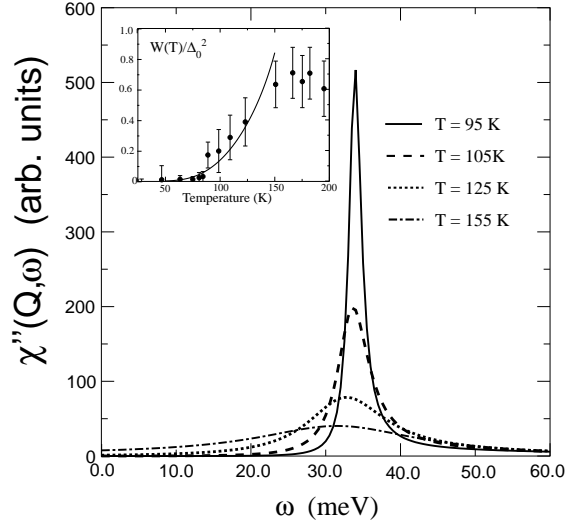}
\end{center}
\caption{The resonance peak in the pseudo-gap region for ${g^2
\alpha \xi^{2} \over 2 v_F^2}=1/(70 meV)$,$\Delta_{sw}=35 meV$,
$\omega_c=65 meV$ and $W(T)$ from the inset. Inset: $W(T)$ from
Ref.~\protect\cite{Fra98} (points) and from Eq.(\ref{Wfinal}) with
A=0.1 and C=0.6 (solid line). } \label{res_peak}
\end{figure}

        Assuming that the superfluid velocity is
purely due to transverse phase fluctuations, we have
\begin{equation}
W(T)=\pi^2 v_F^2  \int
\frac{d^{2}\bf{q}}{4\pi^{2}}\frac{G(\bf{q})}{q^{2}} \ , \label{WM}
\end{equation}
where $ G(\bf{q}) =\left\langle n_{\bf{q}}n_{-\bf{q}}\right
\rangle$ is the vortex density correlator. In the limit of large
vortex density, this correlator is given by \cite{Minnhagen87} $G(
{\bf q})^{-1} \sim 4\pi^2 K_{0}(\xi ^{2}_{\theta }+q^{-2})$.
Evaluation of the integral in Eq.(\ref{WM}) with wave-vector
cutoff $\Lambda = 2 \pi \xi^{-1} _{GL}$ yields
\begin{equation}
W\left( T\right) \simeq  { \pi^3 \Delta^2_{0} \over 8 K_0(T) }
\left( { \xi _{GL} \over 2 \pi \xi _{\theta}} \right)^2 \ln \left[
1+\left({2 \pi\xi _{\theta } \over \xi _{GL}} \right)^{2}\right] \ .
\label{Wfinal}
\end{equation}
Here, the phase coherence length is given by
$\xi_\theta^{-2}(T)=4\pi^2 K_0(T)n_F(T)$, where $n_F(T)$ is the
density of free vortices, and we used the BCS result $\xi
_{GL}=v_F/(\pi\Delta_0)$. Corson {\it et al.}~found in their analysis that
$n_F(T)=(2A/\pi\xi^{2}_{GL})\exp(-8 C K_0(T))$ with $A$ and $C$ being
constants of order $O(1)$. Assuming that the functional form of $n_F(T)$ (and
thus that of $W(T)$) remains the same for all underdoped Bi-2212 compounds,
with $A$ and $C$ being the only doping dependent parameters, we present $W(T)$
from Eq.(\ref{Wfinal}) in the inset of Fig.~\ref{res_peak}, together with FM's
fits \cite{Fra98} to STS experiments in Bi-2212 ($T_c = 83K$) \cite{sts}. With
$A=0.1$ and $C=0.6$, we find good quantitative agreement of our theoretical
results with those of FM up to $T_0 \approx 150$ K. At this temperature, the
above approximation, as well as the analysis by FM, presumably break down
since $\sqrt{W(T)}$ becomes of the order of the maximum
superconducting gap.

We now turn to the appearance of the resonance peak in the
pseudogap region. Morr and Pines (MP) \cite{Morr98} recently
argued that the resonance peak in the superconducting state arises
from a spin-wave mode whose dispersion is given by
\begin{equation}
\omega_q^2=\Delta_{sw}^2+c_{sw}^2|\bf{q}-\bf{Q}|^2 \ ,
\label{omq}
\end{equation}
where $\Delta_{sw}$ is the spin-wave gap, $c_{sw}$ is the
spin-wave velocity and $\bf{Q}$ is the position of the magnetic
peak in momentum space. Starting from a spin-fermion model
\cite{sfmodel}, MP showed that this mode is strongly damped in the
normal state, but becomes only weakly damped in the
superconducting state, if $\Delta_{sw}$ is less than the gap,
$\omega_c$, for particle-hole excitations with total momentum
$\bf{Q}$. These excitations connect points on the Fermi surface
(FS) in the vicinity of $(0,\pi)$ and $(\pi,0)$ (``hot spots"),
and thus for a d-wave gap $ \Delta_{\bf{k}}=\Delta _{0}\, \,
(\cos(k_{x})-\cos (k_{y}))/2 \), $\omega_c \approx 2 \Delta_{0}$.

MP computed $\chi$ using the Dyson-equation
\begin{equation}
\label{chifull} \chi^{-1} =\chi_0^{-1}  - \Pi  \  ,
\end{equation}
where $\chi_0$ is the ``bare" susceptibility and $\Pi$ is the
bosonic self-energy given by the irreducible particle-hole bubble.
For $\chi_0$, MP made the experimentally motivated ansatz
\begin{equation}
\label{chi0} \chi_0^{-1}= { \omega_q^2-\omega^2 \over  \alpha \,
c^{2}_{sw} } \ ,
\end{equation}
where $\omega_q$ is given in Eq.(\ref{omq}). In the
superconducting state, one obtains for $\Pi$ to lowest order in
the spin-fermion coupling $g$
\begin{eqnarray}
\Pi({\bf q}, i \omega_n) &=& - g^2 \, T \sum_{{\bf k},m} \
\Big\{  G({\bf k}, i\Omega_m) G({\bf k+q}, i\Omega_m+i\omega_n) \nonumber  \\
& & + F({\bf k}, i\Omega_m) F({\bf k+q}, i\Omega_m+i\omega_n)  \Big\} \ ,
\label{Pi}
\end{eqnarray}
with $G$ and $F$ being the normal and anomalous Green's functions.
Since, within the spin-fermion model, $\chi_0$ is obtained by
integrating out the high-energy fermionic degrees of freedom, it
is largely unaffected by the onset of superconductivity or the
pseudo-gap. Moreover, MP argued that due to fermionic self-energy
corrections, Re$\, \Pi$ in the SC state only leads to an
irrelevant renormalization of $\Delta_{sw}$ and $c_{sw}$. Since
the same argument also holds within our scenario for the
pseudo-gap region, we neglect Re$\, \Pi$ in the following.

On the other hand, Im$\, \Pi$ which determines the damping of the
spin excitations, changes dramatically in the SC state due to the
opening of a gap in the fermionic dispersion. Consequently, we
expect phase fluctuations to strongly affect Im$\, \Pi$. Moreover,
in each polarization bubble present in the RPA expansion of
Eq.(\ref{chifull}), the electron-hole pairs probe different parts
of the sample and thus independent configurations of thermally
excited super-currents. It then follows that the susceptibility,
$\chi_{pf}$, in the presence of phase fluctuations is obtained
from Eq.(\ref{chifull}) by using Im$\, \Pi_{pf}$ averaged over the
thermodynamic ensemble determined by Eq.(\ref{Hxy}).

\begin{figure} [t]
\begin{center}
\leavevmode \epsfxsize=7.5cm \epsffile{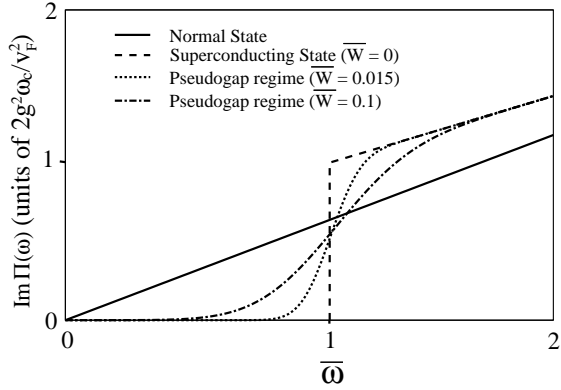}
\end{center}
\caption{${\rm Im}\, \Pi$ in the normal (black line) and SC state
(dashed line), and $ {\rm Im}\, \Pi_{pf} ( \omega)$ in the
pseudo-gap region for {\bf (a)} ${\bar W}=0.015$ (dotted line),
and {\bf (b)} ${\bar W}=0.1$ (dashed-dotted line). }
\label{ImPi}
\end{figure}

Before we discuss the effect of phase fluctuations on Im$\, \Pi$,
we shortly review its form in the normal and SC state. Extending
Eq.(\ref{Pi}) to the normal state, we obtain Im$\, \Pi({\bf
Q})= 4g^2\omega/ (\pi v^{2}_{F})$ \cite{Chu97}, where $v_F$ is the Fermi 
velocity
at the hot spots. In contrast, in the SC state, in the limit of $
T\ll \omega_c $, we find to order $O(T/\omega_c)$
\begin{eqnarray}
{\rm Im}\, \Pi ({\bf Q}, \omega) &=& {4 g^2 \omega_c \over \pi
v_F^2}\,E(\sqrt{{1 - \bar \omega}^2})\, \theta \left( {\bar
\omega}-1 \right) \nonumber \\ &\sim &  {g^2 \omega_c \over v_F^2}
( {\bar \omega} + 1)  \theta \left( {\bar \omega}-1 \right) \ ,
\label{PiBCS}
\end{eqnarray}
where $\theta(x)$ is the Heavyside step function, $E(x)$ is the
complete Elliptic integral of the first kind and  ${\bar
\omega}=\omega/\omega_c$. Thus, Im$\, \Pi$ vanishes for
frequencies below $\omega_c$. In Fig.~\ref{ImPi} we present the
frequency dependence of Im$\, \Pi$ in the normal and SC state.

We now consider the effect of phase fluctuations on Im$\, \Pi$.
Note that $\Pi$, Eq.(\ref{Pi}), and thus $\chi$, Eq.(\ref{chifull}),
are invariant under the gauge transformation,
Eq.(\ref{GT}), in contrast to $G(k,\omega)$, considered by FM.
Thus $G,F$ can be straightforwardly calculated using the
$\Psi$-fermions. In the limit $k_F\xi_\theta\gg 1$, where $k_F$ is the Fermi
momentum at the hot-spots, the interaction of the $\Psi$-fermions with the
superfluid flow leads to a Doppler  shift in the $\Psi$-excitation spectrum
\cite{Fra98,Kwo98} given by
\begin{equation}
\label{qp_disp} E^{\pm }_{\bf{k}}=\sqrt{\epsilon
_{\bf{k}}^{2}+|\Delta _{\bf{k}}|^{2}}\pm D_{\bf{k}} \
\end{equation}
where $\epsilon _{\bf{k}}$ is the fermionic dispersion in the
normal state,  \(D_{\bf{k}}=m \bf{v}_F(\bf{k}) \cdot \bf{v}_{s} \)
is the induced Doppler-shift and $\bf{v}_F(\bf{k})$ is the Fermi
velocity. In the limit $T \ll D_{\bf{k}} \ll \Delta_0$, Im$\, \Pi$
for a given superfluid velocity is obtained from Eq.(\ref{PiBCS})
via the frequency shift
\begin{equation}
\label{om}
\omega \rightarrow \omega +\left( D_x +D_y \right) \ .
\end{equation}
Similar to the case of the fermionic spectral function \cite{Fra98},
the thermodynamic average of Im$\, \Pi$ over the ensemble specified
by Eq.(\ref{Hxy}) is obtained by convoluting Im$\, \Pi$ with a
Gaussian distribution of Doppler shifts of the form
\begin{equation}
\label{prob} P(D_{\alpha })={ 1 \over \sqrt{2\pi W} }
\exp\left( -\frac{D_{\alpha }^2}{2W}\right) \, \, ,
\end{equation}
where \(\alpha =x,y \). In the limit $\sqrt{W} \ll T \ll
\omega_c$,  we can perform this convolution analytically and
obtain
\begin{eqnarray}
{\rm Im}\, \Pi_{pf} (\omega)   & =& {g^2 \omega_c \over 2 v_F^2}
 \Bigg\{ \left( 1+{\bar\omega}\right)
\left[ 1+\Phi \left( \frac{ {\bar\omega} -1}{\sqrt{ {\overline
W}(T)}})\right)\right] \nonumber \\ & + & \sqrt{ {\overline
W}\left( T\right)/\pi } \ \exp\left(-{ ({\bar \omega} -1)^2 \over
 {\overline W}(T) } \right)\Bigg\} \ ,
\label{ImPi_ave}
\end{eqnarray}
where \( \Phi (x) \) is the error function. It follows from
Fig.~\ref{ImPi}, in which we present the spin-damping for two
different values of ${\overline W}=W/\Delta_0^2$, that the effect of phase
fluctuations on Im$\, \Pi$ is two-fold. First, they lead to a non-zero value of
Im$\, \Pi_{pf} ( \omega)$ for $\omega<\omega_c$, in contrast to
the form of Im$\, \Pi$ in the superconducting state where the
spin-damping at $T=0$ vanishes below $\omega_c$. Second, the
spin-damping below $\omega_c$ increases with increasing $W$ while
at the same time, the sharp step in Im$\, \Pi$ is smoothed out.
Note that in the pseudo-gap region, $T\ll \omega_c$, and
consequently the temperature dependence of $ {\rm Im}\, \Pi_{pf} (
\omega) $ is determined by that of $W(T)$.

Finally, inserting ${\rm Im}\, \Pi_{pf} ( \omega) $ into
Eq.(\ref{chifull}), we obtain $\chi_{pf}''({\bf Q}, \omega)$ in
the pseudo-gap region. In Fig.~\ref{res_peak} we present our
theoretical results for the frequency and temperature dependence
of the resonance peak in Bi-2212 ($T_c=83K$), using the $W(T)$
shown on the inset.
We find that, as the temperature is increased above $T_c$, the
resonance peak becomes broader, while its intensity diminishes.
Since $W(T)$ is a monotonically increasing function of
temperature, it follows from Fig.~\ref{ImPi} that the spin damping
for $\omega \approx \Delta_{sw}<\omega_c$ also increases with
temperature, giving rise to the behavior of the peak
intensity/width shown in Fig.~\ref{res_peak}. Unfortunately, no
experimental data for the temperature dependence of  the resonance peaks in
the pseudo-gap region of underdoped Bi-2212 are currently available. However,
our results are in qualitative agreement with the experimental data on
underdoped YBa$_2$Cu$_3$O$_{6+x}$ \cite{Dai}.

\begin{figure} [t]
\begin{center}
\leavevmode
\epsfxsize=7.5cm
\epsffile{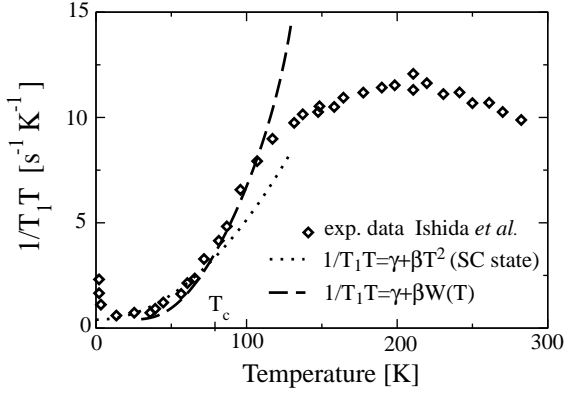}
\end{center}
\caption{$ 1/T_{1}T$ in the pseudo-gap region of underdoped Bi-2212 ($Tc=79K$). 
Solid
line: theoretical fits with A=0.05 and C=0.3. Filled squares:
experimental data taken from Ref.~\protect\cite{Ish98}. We assumed a
constant background factor $\gamma=0.4 K^{-1}s^{-1}$} \label{nmr}
\end{figure}

We now turn to the second experimental probe of $\chi''$, the \(
^{63} \)Cu spin-lattice relaxation rate, \( 1/T_{1} \). For an
applied field parallel to the \( c- \)axis, \( 1/T_{1} \) is given
by
\begin{equation} \label{T1T} \frac{1}{T_{1}T}={k_{B} \over 2}
(\gamma _{n}\gamma _{e})^{2}\frac{1}{N}\sum _{\bf{q}}F_{c}(\bf{q})\lim
_{\omega \rightarrow 0}\frac{\chi'' (\bf{q},\omega )}{\omega }\, \, ,
\end{equation} where \begin{equation}
\label{form} F_{c}({\bf q})=\left[ A_{ab}+2B\left(
cos(q_{x})+cos(q_{y})\right) \right] ^{2}\, \, ,
\end{equation}
 and \( A_{ab} \) and \( B \) are the on-site and transferred
hyperfine coupling constants, respectively. The spin-lattice
relaxation rate in the mixed state, i.e., in the presence of a
superflow, was recently considered by Morr and Wortis (MW)
\cite{Morr99}. Using the low-frequency limit of
Eqs.(\ref{chifull}) and (\ref{Pi}), they found that the
temperature dependence of $1/T_{1}$ is determined by the set
$\{D_n/T\}$, where $D_n$ is the Doppler-shift at the
$n$th node (see Eq.(\ref{qp_disp})). In
the limit, $|D_n/T| \gg 1$, they obtained
\begin{equation}
{1 \over T_1 T} = { {\cal C} \over N} \sum_{i,j} {\cal F}({\bf
q}_{i,j}) |D_i| \, |D_j| \ ,
 \label{T1Tf1}
\end{equation}
where ${\cal C}= (k_B/\pi)(\alpha g \gamma_n
\gamma_e)^2/ (4 v_F v_\Delta)^2$, $v_\Delta=|\partial
\Delta_{\bf k} /\partial {\bf k}|$ at the nodes, and
\begin{equation}
{\cal F}({\bf q}_{i,j})={F_c({\bf q}_{i,j})  \over (\xi^{-2} +
|{\bf q}_{i,j}-{\bf Q}|^2)^2} \ . \label{calF}
\end{equation}
Here, ${\bf q}_{i,j}$ is the wave-vector connecting the nodes $i$
and $j$, and $\xi$ is the magnetic correlation length. In the
limit $T \ll \sqrt{W(T)}$, the convolution of Eq.(\ref{T1Tf1})
with the Gaussian distribution of Eq.(\ref{prob}) can be performed
analytically, and we obtain
\begin{equation}
\left( { 1 \over T_1 T }\right)_{pf} =  \beta \, W(T)
\label{T1T_ave}
\end{equation}
where $\beta=4 {\cal C}({\cal F}(0)+ {\cal F}({\bf q}_{1,3}) +
{8 \over \pi} {\cal F}({\bf q}_{1,2}))$. The constant $\beta$ can be
experimentally obtained \cite{comm1} by fitting $(T_{1}T)^{-1}$ at $T<T_c$
with the d-wave BCS expression $(T_{1}T)^{-1}=\beta{\pi^2 \over 3} T^2$.
Note that the relaxation rate in Eq.(\ref{T1T_ave}) directly
reflects the strength of the classical phase fluctuations. In Fig.~\ref{nmr}
we present our theoretical results for $(T_{1}T)^{-1}_{pf}$,
Eq.(\ref{T1T_ave}), together with the experimental data by Ishida {\it et
al.}~\cite{Ish98} on underdoped Bi-2212 ($T_c=79K$).
Using $W(T)$, Eq.(\ref{Wfinal}), with A=0.05 and C=0.3, we find good
agreement of our theoretical results with the experimental data
between $T_c$ and $T_0 \approx 130$ K.
Note that the external magnetic
field applied in NMR experiments increases the density of free
vortices by ${\bar n}_{F}(B) \sim B/ \phi_{0}$ \cite{Minnhagen87}. As a
result, the phase coherence length (and in turn $W$) acquires a
magnetic field dependence, $\xi ^{-2}_{\theta}=4\pi^2 K_{0}(T)
\left(n_{F}(T)+{\bar n}_{F}(B) \right)$. For $T \geq T_c$, and the
parameter set used above, we find that a magnetic field of $B=10T$
increases the vortex density by ${\bar n}_{F}(B)/ n_{F}(T_c) \sim
0.1$; its contribution to $W$ can thus be neglected. Thus the relaxation rate 
above $T_c$ should be independent of magnetic field for typical values of $B$, 
which is consistent with recent experiments by Gorny {\it et al.} \cite{Gor99}.

In the above scenario, we neglected the effect of longitudinal
phase fluctuations which arise from spin-wave like excitations.
This is justified since their excitation spectrum is very
likely gapped by the Anderson-Higgs mechanism \cite{Fra98}, and
they are, consequently, irrelevant for the low-frequency
thermodynamic properties of the underdoped cuprates. It was recently proposed
in Ref.\cite{Carl99} that longitudinal phase fluctuations are responsible for
the linear temperature dependence of the superfluid density at $T\ll T_c$. FM
pointed out that longitudinal phase fluctuations at $T \ll T_c$
lead to a  $W_{long} \sim T$. In this case it follows from Eq.(\ref{T1T_ave})
that, for $^{63}$Cu and $^{17}$O, $1/T_1T \sim T$ at $T\ll T_c$, in contrast
to the experimentally observed $1/T_1T \sim T^2$ \cite{Mar94}. This result 
suggests
that longitudinal phase fluctuations are absent in the superconducting state.

We assumed above, following the argument applied to STS and ARPES
experiments \cite{Fra98}, that transverse phase fluctuations are
static on the time-scale of INS and NMR experiments which allowed
us to neglect the quantum dynamical nature of the vortices. While
this assumption likely holds for ``fast" probes like INS, ARPES
and STS where the quasi-particles are coupled to phase fluctuations
for short times, it might be less justified for the much
``slower'' NMR experiments. In this light, the agreement of our
theoretical NMR results with the experimental data, Fig.~\ref{nmr}
is remarkable. However, the effects of the vortex quantum dynamics
on various experimental probes is still an open question which
requires further study.

In summary we propose a scenario for INS and NMR experiments in
the pseudogap region of the underdoped cuprates. We argue that
phase fluctuations of the superconducting order parameter
drastically affect the frequency dependence of the spin
susceptibility and can thus qualitatively account for the
temperature dependence of the resonance peak. Moreover, we show
that the spin-lattice relaxation rate, $1/T_1T$, is a direct probe
for the strength of the phase fluctuations, as reflected in
$W(T)$. Finally, we showed that $W(T)$ obtained from high
frequency transport measurements is in good qualitative agreement
with that extracted from STS experiments.

It is our pleasure to thank A. H. Castro Neto, A.V. Chubukov, 
A. J. Leggett, A. J.
Millis, D. Pines, R. Ramazashvili, J. Schmalian, R. Stern and M.
Turlakov for valuable discussions. This work has been supported by
\emph{Funda\c c\~ ao de Amparo \`a Pesquisa do Estado de S\~ ao
Paulo} (FAPESP), the Center of Advanced Studies (CAS) of the
University of Illinois (H.W.) and in part by the Science and
Technology Center for Superconductivity through NSF-grant
DMR91-20000 and by DOE at Los Alamos (D.K.M).


\begin{thebibliography}{99}
\bibitem{comm1} We assume that $\, \Delta_{sw}$ and $c_{sw}$
only exhibit a weak temperature dependence in the
pseudo-gap region \protect\cite{Bourges}.
\end{thebibliography}
\end{document}